\documentclass[aps,twocolumn,pra,showpacs,floatfix]{revtex4-1}
\usepackage{graphicx}
\usepackage{dcolumn}
\usepackage{amsmath}

\begin{document}

\title{Effects of Lorentz Symmetry Violation in the Spectra of Rare-Earth Ions in a Crystal Field}
\author{C. Harabati}
\author{V.A. Dzuba}
\author{V.V. Flambaum}
\affiliation{School of Physics, University of New South Wales,
Sydney 2052, Australia}
\author{M.A. Hohensee}
\affiliation{Lawrence Livermore National Laboratory, Livermore, CA 94550, USA}
\date{\today}

\begin{abstract}
We demonstrate that experiments measuring the transition energies of rare-earth ions doped in crystalline lattices are sensitive to violations of Local Lorentz Invariance and Einstein's Equivalence Principle.  Using the crystal field of LaCl$_{3}$ as an example, we calculate the frame-dependent energy shifts of the transition frequencies between low-lying states of Ce$^{3+}$, Nd$^{3+}$, and Er$^{3+}$ dopants in the context of the Standard Model Extension, and show that they have high sensitivity to electron anomalies that break rotational invariance.    
\end{abstract}

\pacs{03.30.$+$p, 11.30.Er, 11.30.Cp, 32.30.Jc}

\maketitle

Most of our day-to-day experiences are mediated by light and charged particles, and in particular its interaction with electrons.  To the best of our knowledge, the physics of a system of photons and electrons is independent of the velocity and orientation of that system in absolute space, nor is it locally dependent upon where that system lies in a gravitational potential.  These symmetries, respectively described as local Lorentz invariance (LLI) and Einstein's equivalence principle (EEP), are fundamental to our modern understanding of the standard model and general relativity.  It is possible, however, that these symmetries are not exact at experimentally accessible energy scales, thanks to spontaneous symmetry breaking or other physics at high energy scales~\cite{KosteleckySamuel:1989,Damour}.  This possibility has driven many experimental tests of LLI and EEP~\cite{T14}, and motivated the development of phenomenological frameworks that can quantitatively describe the effect of LLI- and EEP-violation on known particles and fields.  One such framework is provided by the standard model extension (SME)~\cite{CK97,CK98}, which has been used to analyze a wide range of experiments~\cite{datatables}.  The SME augments the standard model Lagrangian with all combinations of known particles and fields that are not invariant under Lorentz transformations, but which preserve gauge invariance, energy and translational invariance, and the invariance of the total action~\cite{CK97,CK98}.  These terms are parameterized by Lorentz tensors that are collectively known as LLI- and EEP-violating coefficients, and are further subdivided into 'sectors' that deal with terms involving a particular particle.  In this Letter, we focus on tests of the electron-sector $c_{\mu\nu}$ tensor, which modifies the inertial energy of electrons according to their direction of motion.  

Spectroscopy of neutral dysprosium atoms has already led to one of the world's most sensitive tests of electronic LLI and EEP~\cite{HLBHDF13}.  More recently, a still more sensitive measurement of the electron $c_{\mu\nu}$ coefficients was obtained by engineering the quantum state of a pair of trapped Ca$^{+}$ ions~\cite{PruttivarasinNature}, extending precision tests of electronic LLI past the electroweak (relative to the Planck mass) scale.   Both of these experiments operate at or near the interrogation-time-of-flight or atom (ion) shot-noise limit. In this Letter, we consider the possibility of using rare earth ions doped in a crystalline lattice to perform similar measurements of the electronic $c_{\mu\nu}$.  Rare-earth ion-doped crystals offer substantially larger ion-number densities than are available in atomic gases, and interrogation times comparable and potentially longer than are possible in ion traps.  The relevant 4f orbitals are well screened from one another and from the fluctuations of external fields, yielding the sharp, stable optical transitions that rare earth ion-doped materials are known for at low temperatures.  The strong crystal field produces optical-frequency splitting of the otherwise degenerate free-ion $|J,M\rangle$ states.  This splitting is far larger than could be produced by an externally generated magnetic or electric field, and as a result, is very stable with respect to external field fluctuations.  These properties make rare-earth ion-doped crystals extremely advantageous for fundamental symmetry tests, complementing recent tests focusing on the resonant modes of bulk crystals~\cite{LoSubmitted}.  Our work raises the novel prospect of using solid state systems to test LLI with greater precision than can be achieved by spectroscopy of free particles.

\section{Theory}\label{sec:theory}

In the SME, spin-independent violations of LLI for electrons generate a linearized perturbation of the electron $c_{\mu\nu}$ tensor to the Dirac Hamiltonian which may be written in natural units as~\cite{CK97,CK98,KL99a,KL99b}
\begin{equation}\label{eq:perthamil}
\delta H=-(c_{00}\delta_{jk}+c_{jk})\alpha^j p^k-(c_{0j}+c_{j0})p^j - c_{00}m \beta,
\end{equation}  
where  $c_{\mu\nu}$ is a symmetric, traceless, constant background tensor, $\alpha^j=\gamma^0\gamma^j$ and $\beta=\gamma^0$ are the usual Dirac matrices and $\delta_{ik}$ is the Kronecker symbol.  In general, one can define a coordinate transformation which maps the elements of $c_{\mu\nu}$ to zero, at the cost of generating new LLI-violating terms in the photon (and other matter) Hamiltonians.  Such terms can in turn be constrained by modern Michelson-Morley experiments~\cite{KM02}.  Since our choice of coordinates is arbitrary, it follows that measurements of $c_{\mu\nu}$ are equivalent to Michelson-Morley tests, as both are only sensitive to the differences in LLI-violation in the photon and matter sectors.  At first order, the last two terms of Eq.~\eqref{eq:perthamil} do not contribute to shifts in the transition energies between different electronic bound states, and may be omitted.  We can therefore write Eq.~\eqref{eq:perthamil} in terms of the spherical components  of  irreducible tensor operators as:
\begin{equation}\label{eq:irpertur}
\delta H=-C_0^{(0)}T_0^{(0)} -\sum_{q=-2}^{2}(-1)^q C^{(2)}_q T^{(2)}_{-q}.   
\end{equation} 
In the spherical operator form used in this Letter, the elements of $c_{\mu\nu}$ are written as~\cite{KT11,HMW13}
\begin{align*}
C_{0}^{(0)}&=c_{00}+\tfrac{2}{3}c_{jj}-\tfrac{2U}{3c^2}c_{00} & C_{0}^{(2)}&=\tfrac{1}{6}(3c_{zz}-c_{jj})\\
C_{\pm 1}^{(2)}&=\pm(c_{zx}\pm i c_{zy}) & C_{\pm 2}^{(2)}&=\tfrac{1}{2}(c_{xx}-c_{yy}\pm2ic_{xy})
\end{align*}
where doubled roman indexes indicate a sum over the spatial components of $c_{\mu\nu}$, and $U$ is the local Newtonian gravitational potential.  

The problem of a rare-earth ion in a crystal field has been well studied~\cite{B29,D68,HOS65}.  The total Hamiltonian of a rare-earth ion in a crystal field may be written as $H=H_f+V$, where $H_f$ is the free ion Hamiltonian and $V$ is the electrostatic potential due to the 
crystal environment.  The states of free ions are spherically symmetric, and designated by their total angular momentum $J$ and its projection $M$.  When the ions are inserted into a crystal, the ambient crystal field $V$ breaks spherical symmetry, partially lifting the $(2J+1)$-fold degeneracy of the free ions' energy levels~\cite{B29,D68}.  The crystal potential $V$ can be written as
\begin{equation}\label{eq:cryfield}
V=\sum_k\sum_{q=-k}^{k}B_q^k\mathcal{C}^{(k)}_q,   
\end{equation} 
where $\mathcal{C}^{(k)}_q=\sqrt{4\pi/(2k+1)}Y_{kq}$ are normalized spherical harmonics, and 
$Y_{kq}$ are spherical tensor operators of rank $k$. The summation over $k$ is restricted to even numbers because the contribution of $V$ to leading order energy shifts must come from its even-parity components, and  $k\leq 6$ due to the triangle condition for spherical harmonic integrals (since $l=3$ for rare-earth ions with configuration $4f^N$).  The number of terms in Eq.(\ref{eq:cryfield}) may be further reduced using the discrete point symmetry of the crystal.  The $B^k_q$ coefficients, also known as the crystal field parameters, depend on the structure of the crystal and the electronic wave functions' radial components, and are determined by a least-squares fit to the experimental energy levels of the ion in the crystal.  The crystal field potential is  assumed  to act only on the electrons in an open shell, i.e. $4f^N$ for rare-earth ions.  

We obtain the crystal field-induced energy splitting from the secular determinant $|\langle JM|V|J'M'\rangle-\lambda \delta_{JJ'}\delta_{MM'}|=0$ acting on the free ion states $|JM\rangle$.  Diagonalization of $V$ separately within each $J$-manifold yields the split eigenstates $|\psi\rangle$, so that $|\psi\rangle=\sum_{M} a_{M}|JM\rangle$.  Thus the non-zero matrix elements of \eqref{eq:irpertur} are restricted to $\langle JM|T^{2}_{q}|JM'\rangle$, with $q=M-M'$, and $|q|\leq 2$.  

The LLI-violating correction $\delta\omega_{nm}=(\delta E_{n}-\delta E_{m})/\hbar$ to the transition frequency $\omega_{nm}$ between each pair of levels $n$ and $m$ of the ion in the crystal field is a linear combination of the spatial components of the $c_{\mu\nu}$ tensor, which is itself a frame-dependent quantity.  This energy shift also varies as a function of the ion's position in an external gravitational field, although as we will see, this effect is smaller than the frame-dependent phenomena for the transitions of interest.  To uniquely specify the value of $c_{\mu\nu}$, we must also specify the inertial frame in which it is defined.  This frame is typically taken to be approximated by the rest frame of the Sun: specifically the sun-centered celestial equatorial frame (SCCEF), denoted by coordinates $(T,X,Y,Z)$, while the local laboratory frame coordinates are denoted as $(t,x,y,z)$.  For a terrestrial laboratory, the lab-frame values of the tensor's dominant spatial components $c_{jk}^{\rm lab}$ depend upon the orientation of the lab with respect to the SCCEF, and thus modulate with characteristic frequency $\Omega\simeq 2\times2\pi/$(23 h 56 min), or twice every sidereal day.  The value of the anomalous tensor in the lab-frame can be related to that in the SCCEF via $c^{\rm lab}_{\mu\nu}=\Lambda_{\mu}^{\phantom{\mu}\alpha}\Lambda_{\nu}^{\phantom{\nu}\beta}c^{\rm SCCEF}_{\alpha\beta}$, where  $\Lambda_{\mu}^{\phantom{\mu}\alpha}$ is the standard Lorentz boost plus rotation from the SCCEF to the lab frame~\cite{KL99a,BKLR03,HLBHDF13}.  Thanks to the Earth's orbital velocity, the boost $\Lambda_{\mu}^{\phantom{\mu}\alpha}$ mixes the time and spatial components of $c^{\rm SCCEF}_{\alpha\beta}$ into the spatial components of $c^{\rm lab}_{\mu\nu}$.  This gives a measurement that searches for yearly modulations of $\delta H$ access to the parity-odd $c^{\rm SCCEF}_{TJ}$ and the isotropic $c^{\rm SCCEF}_{TT}$ components of the anomalous tensor, albeit with a sensitivity that is suppressed by one and two factors of the Earth's orbital boost velocity $\beta_\oplus\simeq1\times 10^{-4}$.  In what follows, we will focus on the laboratory-frame values of $c_{\mu\nu}$ and the corresponding spherical operator elements $C^{(0)}_0$ and $C^{(2)}_q$, and drop the frame-identifying superscript.

\section{Results and discussion}
\subsection{One valence electron: Ce$^{3+}$ ion in LaCl$_3$\label{subsec:ce}}
Given the available  eigenstates of the rare-earth ions in the crystal field, we can easily calculate the perturbation $\delta\omega_{nm}$ due to Eq.~\eqref{eq:irpertur}.  For trivalent rare earth ions ($R^{3+}$) in the LaCl$_{3}$ lattice (see, e.g., ~\cite{D68}, pg. 149), the crystal field has the point symmetry $C_{3h}$, and is determined by four crystal field parameters $B_{q}^{k}$:
\begin{equation}\label{eq:crypot}
V=B_0^2\mathcal{C}_0^{(2)}+B_0^4\mathcal{C}_0^{(4)}+B_0^6\mathcal{C}_0^{(6)}+B_6^6(\mathcal{C}_6^{(6)}+\mathcal{C}_{-6}^{(6)}).
\end{equation}
The simplest rare earth ion to which Eq.~\eqref{eq:crypot} applies is Ce$^{3+}$, with configuration $4f^1$.  Following the labeling and methods of~\cite{HOS65}, the eigenstates and corresponding energies of Ce$^{3+}$ are presented in Table \ref{t:eigenv}. In the first column of the table, labels of the states are taken from the Ref.  \cite{HOS65}. The eigenstates are also distinguished by their crystal quantum number $\mu$.

The states of the single valence electron $4f^1$ of Ce$^{3+}$ are linear combination of Dirac spinors with mixing of free ion states with angular momenta $J=7/2$ and $J=5/2$. 
These levels' LLI-violating energy shifts~(\ref{eq:irpertur}) follow from the Wigner-Eckart theorem and  the reduced matrix element of the tensor $\mathcal{C}^{(2)}_q$. In terms of the expansion coefficients and radial integrals $I(\kappa',\kappa)$ given in the Appendix, the states' shifts are linear combinations of radial integrals  $I(3,3)=-50.0$, $I(-4,-4)=-49.33$, and $I(-4,3)=-49.58$ in atomic units (a.u.). Here the radial integrals can be obtained from the formulae $I_5$ and $I_1$ in the Appendix, and are taken over Hartree-Fock (HF) wave functions. The total shift of each state is presented in the rightmost column of Table \ref{t:eigenv}. The largest relative energy shift is that between the ground $|I\rangle$ state and the low-lying $|II\rangle$ state, with $\delta\omega_{I,II}=(2\pi) C_{0}^{(2)}(2.76\times 10^{16} {\rm Hz})$.  A similarly large LLI-violating energy shift is observed for the $|I\rangle$ to $|b\rangle$ transition. Details of this calculation are presented in the Appendix.

\begin{table*}
\caption{Experimental crystal-field splittings, calculated wave functions, and LLI-violating energy shifts $\delta E$  for Ce$^{3+}$ ($4f^1$) ions in LaCl$_3$.  Data from reference \cite{HOS65}. $B_0^2=129$, $B_0^4=-329$, $B_0^6=-997$,
 and $B_6^6=403$ in cm$^{-1}$ are used in calculating Eq.~(\ref{eq:crypot}).   Note that the eigenstates are doubly degenerate in non-magnetic crystals.}
\label{t:eigenv}
\begin{ruledtabular}
  \begin{tabular}{l l l l l c}
  Config. & State & $\mu$ & Wave functions $|\psi_\mu\rangle$ & $E$ (cm$^{-1}$) & $\delta E$ ($C_0^{(2)}\times 10^6$cm$^{-1}$)\\
\hline
$^2$F$_{7/2}$ &$|d\rangle$ & $\pm 3/2$ & $0.99924|7/2,\mp 3/2\rangle\pm0.03905|5/2,\mp 3/2\rangle$ & 2399.5 &  1.54\\ 
&$|c\rangle$ & $\pm 1/2$ & $\pm0.82174|7/2,\pm 7/2\rangle\pm0.56692|7/2,\mp 5/2\rangle+0.05785 |5/2,\mp 5/2\rangle$ & 2282.6 &  -2.52\\
&$|b\rangle$ & $\pm 5/2$  & $\mp0.99446|7/2,\mp 1/2\rangle+0.10511|5/2,\mp 1/2\rangle$ & 2208.6 &  2.47\\
&$|a\rangle$ & $\pm 1/2$ & $\pm0.56659|7/2,\pm 7/2\rangle\mp0.82356|7/2,\mp 5/2\rangle+0.02709|5/2,\mp 5/2\rangle$ & 2166.0 & -1.56\\
\hline
$^2$F$_{5/2}$ &$|III\rangle$ & $\pm 3/2$ & $0.03905|7/2,\mp 3/2\rangle\mp0.99924|5/2,\mp 3/2\rangle$ & 110.0 &  0.62\\ 
&$|II\rangle$ & $\pm 5/2$ & $\pm0.10511|7/2,\mp 1/2\rangle+0.99446|5/2,\mp 1/2\rangle$ & 37.5 &  2.61\\
&$|I\rangle$ & $\pm 1/2$  & $\mp0.06298|7/2,\pm 7/2\rangle\mp0.01072|7/2,\mp 5/2\rangle+0.99796|5/2,\mp 5/2\rangle$ & 0.0 & -3.16  
\end{tabular}
\end{ruledtabular}
\end{table*}

In contrast to the case of neutral dysprosium, the contribution of the scalar $T_{0}^{(0)}$ component of the LLI-violating perturbation in Eq.~\eqref{eq:irpertur} to the ions' transition energies is smaller than that of the tensor operator $T_{q}^{(2)}$~\cite{HLBHDF13}.  This occurs because the low-lying ion excitations are largely between states with the same quantum number $n$ and total angular momentum $J$.  In the non-relativistic limit, the scalar operator $T_{0}^{(0)}$ is proportional to the sum of the bound electrons' kinetic energy $\sum_i{\bf p}_i^2/2m$.  Using the virial theorem in a Coulomb potential, the electrons' binding energy is approximately equal to their kinetic energy.  The LLI-violating change in the transition energy between bound states with the same $n$ and different $J$ is therefore expected to scale as $\delta \omega_{nm}\simeq C_{0}^{(0)}(6\times 10^{13} {\rm Hz})$.  This, combined with the comparatively small range of variation in $U/c^2$ accessible to a terrestrial laboratory ($\sim 10^{-10}$ in the Sun's potential over a year), implies this ion could serve as a stable reference standard to compare against a more sensitive transition (such as those offered by dysprosium~\cite{HLBHDF13}) in a null-redshift test of EEP.

\begin{table*}
\caption{Experimental crystal-field splittings, calculated wave functions, and LLI-violating energy shifts $\delta E$ for the ground levels Nd$^{3+}$ ($4f^3$, $^4I_{9/2}$) and Er$^{3+}$ ($4f^{11}$, $^4I_{15/2}$) in LaCl$_3$. Data from Ref. \cite{GS74}, in terms of crystal quantum numbers $\mu$ from Ref.~\cite{W65}. }
\label{t:manyelec}
\begin{ruledtabular}
  \begin{tabular}{l l l l l c }
 Species & State & $\mu$ & Wave functions $|\psi_n\rangle$ & $E$ (cm$^{-1}$) & $\delta E$ ($C_0^{(2)}\times 10^6$cm$^{-1}$)\\
\hline
Nd$^{3+}$ &$|e\rangle$ & $\mp 3/2$ & $0.558|\pm 9/2\rangle-0.830|\mp 3/2\rangle$ & 249.4 &  0.05\\ 
&$|d\rangle$ & $\pm 5/2$ & $0.936|\pm 5/2\rangle-0.351|\mp 7/2\rangle$ & 244.4 &  0.16\\
&$|c\rangle$ & $\mp 3/2$  & $0.830|\pm 9/2\rangle+0.558|\mp 3/2\rangle$ & 123.2 &  -0.83\\
&$|b\rangle$ & $\pm 1/2$ & $|\pm 1/2\rangle$ & 115.4 &  1.03\\
&$|a\rangle$ & $\pm 5/2$ & $0.351|\pm 5/2\rangle+0.936|\mp 7/2\rangle$ & 0.0   &  -0.42\\
\hline
Er$^{3+}$&$|h\rangle$ & $\pm 1/2$ & $0.905|\pm 13/2\rangle+0.356|\pm 1/2\rangle+0.232|\mp 11/2\rangle$ & 229.31 &  1.31\\ 
&$|g\rangle$ & $\mp 1/2$ & $0.820|\pm 11/2\rangle +0.429|\mp 1/2\rangle -0.379|\mp 13/2\rangle$ & 181.04 &  0.45\\
&$|f\rangle$ & $\pm 3/2$ & $0.116|\pm 15/2\rangle+0.764|\pm 3/2\rangle+0.635|\mp 9/2\rangle$  & 141.61 & -0.96 \\
&$|e\rangle$ & $\mp 5/2$ & $0.662|\pm 7/2\rangle+0.750|\mp 5/2\rangle$ & 113.7  &  -1.08\\
&$|d\rangle$ & $\pm 1/2$ & $0.192|\pm 13/2\rangle-0.830|\pm 1/2\rangle+0.523|\mp 11/2\rangle$ & 96.52  &  -0.26\\ 
&$|c\rangle$ & $\pm 3/2$ & $0.925\pm 15/2\rangle+0.150|\pm 3/2\rangle-0.349|\mp 9/2\rangle$ & 64.27  &  2.58\\
&$|b\rangle$ & $\pm 3/2$ & $0.362|\pm 15/2\rangle-0.628|\pm 3/2\rangle+0.689|\mp 9/2\rangle$  & 37.91  &  -0.29\\
&$|a\rangle$ & $\mp 5/2$  & $0.750|\pm 7/2\rangle-0.662|\mp 5/2\rangle$ &  0.0    &  -1.02
\end{tabular}
\end{ruledtabular}
\end{table*}
\subsection{Several valence electrons: Nd$^{3+}$ and Er$^{3+}$ ions in LaCl$_3$.}
We can perform a similar calculation for rare earth ions with several valence electrons in LaCl$_{3}$.  Using the energy levels and approximate wave functions for Nd$^{3+}$ and Er$^{3+}$ available in the literature~\cite{GS74}, we find that these ions have respective ground state configurations ($4f^3$, $^4I_{9/2}$) and ($4f^{11}$, $^4I_{15/2}$).  In this case, the relevant electronic wave functions are linear combinations of Slater determinants of HF orbitals.  As before, the fine structure manifolds of these ions are split by a crystal field with point symmetry $C_{3h}$, and as before, the contribution of the $C_{0}^{(0)}$ anomaly to the observed transition frequencies is much smaller than that of $C_{0}^{(2)}$.  Repeating the analysis of part~\ref{subsec:ce}, we may write
\begin{widetext}
\begin{eqnarray}\label{eq:melectron}
\delta E(J,\mu)=-C^{(2)}_0\langle \mu J \parallel T^{(2)}\parallel\mu J \rangle \frac{ 3\sum_M a_M^2 M^2-J(J+1)}{\sqrt{(2J+3)(J+1)(2J+1)J(2J-1)}} 
\end{eqnarray}
\end{widetext}
where $a_M$ are the coefficients for the wave functions presented in Tab.~\ref{t:manyelec}.  Inspection of Tab.~\ref{t:manyelec} reveals that the maximal change in the ions' transition frequencies due to LLI-violation is $\delta \omega_{b,c}=2\pi C_0^{(2)}(5.6\times10^{16}{ \rm Hz})$ for Nd$^{3+}$ and $\delta\omega_{c,e}=2\pi C_{0}^{(2)}(11\times 10^{16} {\rm Hz})$ for Er$^{3+}$.  As in the case of Ce$^{3+}$, the scalar shift proportional to $C_0^{(0)}$ is expected to be comparatively negligible.

A dedicated experiment measuring the THz-scale energy splitting between the $|c\rangle$ and $|e\rangle$ states of Er$^{3+}$ at the level of 1 mHz would be sensitive to $C^{(2)}_{0}$ as small as $10^{-20}$.  To estimate the reach of existing experimental measurements of Er$^{3+}$ transitions, we have also considered optical transitions in Er$^{3+}$:Y$_2$SiO$_5$ \cite{BSTC06}.  Because of the lower symmetry $C^6_{2h}$ of the crystal Y$_2$SiO$_5$ field, each $J$ manifold is split into $J+1/2$ doubly degenerate states~\cite{W65}.  We focus specifically on the $Z_{1}\rightarrow Y_{1}$ transition between the lowest energy levels in the ground $^{4}I_{15/2}$ and excited $^{4}I_{13/2}$ manifolds~\cite{BSTC06}.  Using Eq.~\eqref{eq:melectron}, the LLI-violating perturbation proportional from $C^{(2)}_{0}$ to this transition's energy can be obtained from the reduced matrix elements: $\langle \mu J \parallel T^{(2)}\parallel\mu J \rangle =-67.097$ a.u. for $J=15/2$ and $-56.188$ a.u. for $J=13/2$.  As for the case of the lowest level of Er$^{3+}$:LaCl$_{3}$, we have taken the weighted sum $\sum_M a_M^2 M^2\approx 10$ for  both levels $Z_1$ and $Y_1$ (see Table~\ref{t:manyelec}).  This yields the frequency shift $\delta \omega_{Z_1Y_1}=C_{0}^{(2)}(1.03\times 10^{16} {\rm Hz})$.  Though not considered here, we note the $C^{(2)}_{\pm2}$ component of the LLI-violating tensor might also contribute to $\delta\omega_{Z_{1}Y_{1}}$.
\subsection{Magnetic field effect}
Stray magnetic fields can produce major systematic errors in Lorentz symmetry tests.  We have therefore estimated the effects of magnetic fields on such tests which use Ce$^{3+}$ and Nd$^{3+}$ in LaCl$_3$ crystals.  Detailed studies of the Zeeman effect in Ce$^{3+}$ ions may be found in reference~\cite{BV69}, wherein the first order Zeeman splitting of doubly degenerate states may be found.  Though the ground state degeneracy is lifted by the $2.1\mu_{\rm B}$ (where $\mu_{\rm B}=0.467$cm$^{-1}/$T is the Bohr magneton), the magnetic interaction between states in different doublets is restricted due to the crystal symmetry and the selection rules for M1 transitions.  For Ce$^{3+}$, the quadratic shift has been estimated using the dominant components of the ground state $|I\rangle$ and the low lying state $|III\rangle$  at 110 cm$^{-1}$ (see Tab.~\ref{t:eigenv}).  Since the magnetic quantum numbers differ for these states, transitions are only possible by way of the $x$ and $y$ components of the magnetic moment operator.  Contributions from other levels corresponding to the $^2$F$_{7/2}$ term are suppressed due to the larger energy splittings.  Thus we estimate the second order shift to be $0.918/(110\text{cm}^{-1})\mu_{\rm B}^2$.  This corresponds to a quadratic shift of $5.46\times 10^7$ Hz/T$^2$ for the ground state.  The same calculation can be done for the $|II\rangle$ state, for which we obtain $13.2\times 10^7$ Hz/T$^2$.

The Nd$^{3+}$ ion has three valence electrons in the $f$ shell, and so we limit ourselves to a rough estimate, which may also be applied to other rare-earth ions.  The total angular momentum $J=9/2$ is bigger than the total spin $S=3/2$, so the magnetic moment of the ion is dominated by the angular momentum operator ${\bf J}$.  Hence the matrix elements for transitions within a given multiplet are easily obtained.  As a result, the quadratic shifts are expected to be $\sim 30\times 10^7$ Hz/T$^2$ for the ground level  $|a\rangle$ and $\sim 18\times 10^8$ Hz/T$^2$ for the first excited doublet $|b\rangle$ (see Tab.~\ref{t:manyelec}).

Using these estimates, we may determine the extent to which magnetic fields must be controlled to suppress their effects to below that of an LLI-violating $C_0^{(2)}$ with order $10^{-20}$.  For Ce$^{3+}$ ions, the magnetic field must be stabilized to be below $4.7\times 10^{-6}$ T, or $47$ mG, while for Nd$^{3+}$ ions, a field of no more than $15$ mG should be sufficient.  We further note that larger magnetic field fluctuations, and larger DC fields, are in principle tolerable for a test of LLI, so long as they fluctuate on timescales that are sufficiently different from the modulation periods of the laboratory's orientation and boost relative to a fixed inertial frame (\emph{e.g.} as the SCCEF approximates).
\section{Conclusion}
We have demonstrated that solid state systems, and particularly the ground state spectrum of rare earth ions doped in a crystalline host can be used to perform sensitive tests of LLI. We have taken advantage of existing work on the spectrum of Ce$^{3+}$~\cite{HOS65}, Nd$^{3+}$, and Er$^{3+}$~\cite{E63a,E63b} to perform an explicit calculation of these ions' energy shifts in response to LLI-violation when doped in LaCl$_3$.  The energy levels and sensitivities of Ce$^{3+}$, Nd$^{3+}$, and Er$^{3+}$ ions are expected to be similar when doped in different crystalline media.  Er$^{3+}$ is a particularly interesting case, as the optical coherence of the $J=15/2\rightarrow J=13/2$ transition is particularly long-lived at 4.4 ms for $0.001\%$ doping concentration Er$^{3+}$:Y$_{2}$SiO$_{5}$~\cite{ThielBottgerCone2011}.  An experiment that is sensitive to a 1 mHz orientation-dependent modulation of the $Z_{1}\rightarrow Y_{1}$ optical transition in Er$^{3+}$:Y$_{2}$SiO$_{5}$ could measure spatial components of the electronic $c_{\mu\nu}$ tensor as small as $10^{-19}$, improving upon existing limits by an order of magnitude~\cite{PruttivarasinNature,HLBHDF13}.  An experiment measuring orientation-dependent modulations of the THz-scale energy difference between the $|c\rangle$ and $|e\rangle$ states (see Tab.~\ref{t:manyelec}) at the mHz would be more sensitive still, probing the LLI-violating $c_{\mu\nu}$ tensor at the level of $10^{-20}$.  We note that dynamic decoupling techniques~\cite{ZHABBWLS15}, which switch between states with quantum numbers of equal magnitude and opposite sign do not suppress the LLI-violating signal proportional to the quadrupole component of $C_{0}^{(2)}$ of $c_{\mu\nu}$.  Other rare earth ion-doped materials may also prove to be useful for testing this and other aspects of LLI, and are a promising area for future work.

\begin{acknowledgements}

The authors are grateful to Dmitry Budker and Nathan Leefer  for stimulating discussions.
The work was funded by the Australian Research Council.
V.V. Flambaum is grateful to Humboldt foundation for support and MBN Research Center for hospitality.  This work was performed under the auspices of the U.S. Department of Energy by Lawrence Livermore National Laboratory under Contract DE-AC52-07NA27344.
\end{acknowledgements}

\section*{APPENDIX}
In what follows, we restrict the crystal quantum number to $-\mu$ because  both degenerate states get the same shift from the LLI-violating perturbation. The eigenstates $|I\rangle$, $|a\rangle$, and $|c\rangle$ are presented in Ref. \cite{HOS65}:
\begin{equation}\label{eq:ket1}
\vert\psi_{-\mu} \rangle = \xi\vert 7/2,-7/2\rangle +\eta\vert 7/2,5/2\rangle +\zeta\vert 5/2,5/2\rangle,
\end{equation}
with expansion coefficients $\xi$, $\eta$ and $\zeta$.  Similarly, the remaining eigenstates $|II\rangle$, $|III\rangle$, $|b\rangle$, and $|d\rangle$ are:
\begin{equation}\label{eq:ket2}
\vert\psi_{-\mu} \rangle = \xi\vert 7/2,m\rangle +\zeta\vert 5/2,m\rangle,
\end{equation}
where $m=1/2$ or $3/2$. These basis states are approximated by the relativistic, four-component spinor Hartree-Fock (HF) orbitals of the free Ce$^{3+}$ ion. 
\begin{eqnarray}\label{eq:diracspi}
\psi_{n\kappa m}({\bf r}) = \frac{1}{r}\left(   { f_{n\kappa}(r)\Omega_{\kappa m}(\theta,\phi) \atop i\alpha g_{n\kappa}(r)\Omega_{-\kappa m}(\theta,\phi)}\right),
\end{eqnarray}
where the non-relativistic two-component spinor is defined by
\begin{eqnarray}\label{one-half}
\Omega_{\kappa m}(\theta , \phi)  = \left(  {\pm \sqrt{\frac{\kappa+1/2-m }{2\kappa+1} } Y_{l,m-1/2}(\theta, \phi) \atop \sqrt{\frac{\kappa+1/2+m }{2\kappa+1} } Y_{l,m+1/2}(\theta, \phi)}\right),
\end{eqnarray}
and $\kappa=\mp (j+1/2)$ (for $j=l\pm 1/2$) is the unified quantum number denoting angular momentum and parity.

For transitions between states with the same quantum number $n$ and total angular momentum $J$, the main contribution of the LLI-violating perturbation comes from the tensor $T_{q}^{(2)}$ operator.  Its matrix elements may be written as
\begin{eqnarray}\label{eq:we}
\langle n'\kappa' m'|T^{(2)}_q|n\kappa m\rangle & = & (-1)^{j'-m'}\left(  {j' \atop -m'}{2 \atop q}
{j \atop m}\right) \nonumber\\
 &\times & \langle \kappa' \parallel \mathcal{C}^{(2)}_{q}\parallel \kappa \rangle I(\kappa', \kappa),
\end{eqnarray}
where $I(\kappa', \kappa)$ represents a radial integral~\cite{S07}. 
The reduced matrix element of the tensor $\mathcal{C}^{(2)}_q$ is
\begin{eqnarray}\label{eq:mec2}
\langle \kappa' \parallel \mathcal{C}^{(k)}_{q}\parallel \kappa \rangle=(-1)^{j'+\frac{1}{2}}\sqrt{[j',j]}\left(  {j' \atop -\frac{1}{2}}{j \atop \frac{1}{2}}
{k \atop 0}\right),
\end{eqnarray} 
where $[j',j]\equiv(2j'+1)(2j+1)$.
The energy shifts due to LLI violation can be calculated using equations (\ref{eq:we}), and (\ref{eq:mec2}). In terms of the expansion coefficients and radial integrals $I(\kappa',\kappa)$, the shifts of the states  
$|I\rangle$, $|a\rangle$, and $|c\rangle$ in  Eq. ~(\ref{eq:ket1}) are 
\begin{eqnarray}\label{eq:shift1ac}
\delta E = \frac{C_0^{(2)}}{21}( (7\xi^2+\eta^2)I(-4,-4)+\nonumber\\
+6\zeta^2 I(3,3)+2\sqrt{6}\zeta\eta I(-4,3)).
\end{eqnarray}
Two other states $|II\rangle$ and $|b\rangle$ have the shifts 
\begin{eqnarray}\label{eq:shift2b}
\delta E = -\frac{C_0^{(2)}}{21}( 5\xi^2I(-4,-4)
+\frac{24}{5}\zeta^2 I(3,3)-\nonumber\\
-\frac{4\sqrt{6}}{5}\zeta\xi I(-4,3))
\end{eqnarray}
with $m=1/2$ in Eq.~(\ref{eq:ket2}), while the states $|III\rangle$ and $|d\rangle$ have the energy shifts of the levels 
\begin{eqnarray}\label{eq:shift3d}
\delta E = -\frac{C_0^{(2)}}{21}( 3\xi^2I(-4,-4)
+\frac{6}{5}\zeta^2 I(3,3)-\nonumber\\
-\frac{6\sqrt{2}}{\sqrt{5}}\zeta\xi I(-4,3))
\end{eqnarray}
Here the radial integrals can be obtained from the formulae $I_5$ and $I_1$ below, and are taken over HF wave functions. 

The radial integrals included in Eq.~(\ref{eq:shift1ac},\ref{eq:shift2b},\ref{eq:shift3d}) are from Ref. \cite{HLBHDF13}, and are summarized in this Appendix for convenience. According to the quantum numbers $\kappa'$ and $\kappa$, the radial integrals take  one of the following forms:
\begin{eqnarray}
I_1=c \alpha\hbar  \int_0^\infty\text{d}r
\biggl( (2\kappa -1)g_{n'\kappa'}\frac{\partial f_{n\kappa}}{\partial r}+(2\kappa +3)f_{n'\kappa'}\frac{\partial g_{n\kappa}}{\partial r}\nonumber \\
-\frac{(2\kappa-1)(\kappa +1)}{r}g_{n'\kappa'}f_{n\kappa}-\frac{(2\kappa+3)\kappa}{r}f_{n'\kappa'}g_{n\kappa} \biggr)\nonumber
\end{eqnarray}      
for $\kappa'=-\kappa-1$,  
\begin{eqnarray}
I_2=c \alpha\hbar  \int_0^\infty\text{d}r
\biggl( (2\kappa -3)g_{n'\kappa'}\frac{\partial f_{n\kappa}}{\partial r}+(2\kappa +1)f_{n'\kappa'}\frac{\partial g_{n\kappa}}{\partial r}\nonumber \\
+\frac{(2\kappa-3)\kappa}{r}g_{n'\kappa'}f_{n\kappa}+\frac{(2\kappa+1)(\kappa-1)}{r}f_{n'\kappa'}g_{n\kappa} \biggr)\nonumber
\end{eqnarray}      
for $\kappa'=-\kappa+1$,  
\begin{eqnarray}
I_3=4c \alpha\hbar  \int_0^\infty\text{d}r
\biggl( f_{n'\kappa'}\frac{\partial g_{n\kappa}}{\partial r}
+\frac{\kappa - 1}{r}f_{n'\kappa'}g_{n\kappa} \biggr)\nonumber
\end{eqnarray}      
for $\kappa'=\kappa-2$,  
\begin{eqnarray}
I_4=-4c \alpha \hbar \int_0^\infty\text{d}r
\biggl( g_{n'\kappa'}\frac{\partial f_{n\kappa}}{\partial r}
-\frac{\kappa + 1}{r}g_{n'\kappa'}f_{n\kappa} \biggr)\nonumber
\end{eqnarray}  
for $\kappa'=\kappa+2$, and  
\begin{eqnarray}
I_5=-2c \alpha \hbar \int_0^\infty\text{d}r
\biggl( g_{n'\kappa'}\frac{\partial f_{n\kappa}}{\partial r}-f_{n'\kappa'}\frac{\partial g_{n\kappa}}{\partial r}\nonumber \\
+\frac{\kappa}{r}g_{n'\kappa'}f_{n\kappa}+\frac{\kappa}{r}f_{n'\kappa'}g_{n\kappa} \biggr)\nonumber
\end{eqnarray}      
for $\kappa'=\kappa$.    
In the nonrelativistic limit the matrix element $\langle\phi |c\alpha^j p^k| \phi\rangle$ becomes $\langle \phi|p^jp^k/m|\phi\rangle$. This changes the radial integrals in Eq.~(9). Only the radial integrals differ between the relativistic and the nonrelativistic limits, while the angular parts of the matrix elements are the same. In the non-relativistic limit, $I_1$ and $I_5$ reduce to  
\begin{eqnarray}\label{p2me}
I_1=\hbar^2\int_0^\infty\text{d}r
\biggl( \frac{\partial f_{n'\kappa '}}{\partial r}
\frac{\partial f_{n\kappa}}{\partial r}
+\frac{\kappa(\kappa + 1)}{r^2}f_{n'\kappa'}f_{n\kappa}\biggr)\nonumber
\end{eqnarray} 
$I_2$ and $I_3$ become
\begin{eqnarray}
I_2=\hbar^2\int_0^\infty\text{d}r
\biggl( \frac{\partial f_{n'\kappa '}}{\partial r}
\frac{\partial f_{n\kappa}}{\partial r}-\frac{2\kappa - 1}{r}f_{n'\kappa'}\frac{\partial f_{n\kappa}}{\partial r}\nonumber \\
-\frac{\kappa(\kappa - 2)}{r^2}f_{n'\kappa'}f_{n\kappa}\biggr)\nonumber
\end{eqnarray}    
and $I_4$ takes the form of
\begin{eqnarray}
I_3=\hbar^2\int_0^\infty\text{d}r
\biggl( \frac{\partial f_{n'\kappa '}}{\partial r}
\frac{\partial f_{n\kappa}}{\partial r}+\frac{2\kappa +3}{r}f_{n'\kappa'}\frac{\partial f_{n\kappa}}{\partial r}\nonumber \\
-\frac{(\kappa+3)(\kappa +1)}{r^2}f_{n'\kappa'}f_{n\kappa}\biggr).\nonumber
\end{eqnarray}

\end{document}